\documentclass{article}
\pdfoutput=1

 \usepackage[final,nonatbib]{neurips_2019}

\usepackage[utf8]{inputenc} 
\usepackage[T1]{fontenc}    
\usepackage{hyperref}       
\usepackage{url}            
\usepackage{booktabs}       
\usepackage{amsfonts}       
\usepackage{nicefrac}       
\usepackage{microtype}      

\usepackage{graphicx}
\usepackage{subcaption} 
\usepackage[font=small,skip=0pt]{caption}
\usepackage{amsmath}
\usepackage{amssymb}
\usepackage[]{algorithm2e}

\title{Machine learning based co-creative design framework}

\author{%
  Brian Quanz, Wei Sun, Ajay Deshpande, Dhruv Shah, Jae-eun Park \\
  IBM Research \\
  \texttt{\{blquanz,sunw,ajayd\}@us.ibm.com} \\
}

\begin{document}

\maketitle

\begin{abstract}
 We propose a flexible, co-creative framework bringing together multiple machine learning techniques to assist human users to efficiently produce effective creative designs. We demonstrate its potential with a perfume bottle design case study, including human evaluation and quantitative and qualitative analyses. \end{abstract}

Computers can assist creative processes in many ways, from facilitating idea representation and manipulation to directly collaborating through evaluating, generating, and refining ideas \cite{lubart2005can}.  
Directly collaborating systems are referred to as co-creative or mixed-initiative co-creative (MI-CC) systems, as both the human and computer proactively and creatively contribute to a solution (\textit{e.g.}, a design) \cite{novick1997mixed,liapis2016can,yannakakis2014mixed}, with a goal of inspiring and facilitating lateral thinking, to surpass individual creative outcomes \cite{liapis2016can}.  Most prior MI-CC work focuses on specific, singular applications (\textit{e.g.}, largely video game level design) and single creative-assistant elements (\textit{e.g.}, only suggesting new candidate designs) \cite{liapis2016can,liapis2013sentient,liapis2016mixed,deterding2017mixed,guzdial2016game,guzdial2019friend,guzdial2018co,gilon2018analogy,siangliulue2016ideahound}, which cannot address the diversity of user needs \cite{lubart2005can} and may limit potential and adaption for wider use.  Additionally most past work relies on heuristics and evolutionary or random constrained search  \cite{liapis2016can,liapis2013enhancements,singh2012towards,kim2000application,buelow2008}.  Only recently has machine learning (ML) been incorporated in MI-CC, in limited uses of either learning user preference \cite{liapis2013designer} or learning to generate content (designs) \cite{guzdial2018co}; however, we argue ML has much greater potential to assist co-creative design.

To address these shortcomings, we propose a general co-creative framework, showing how multiple ML methods can be incorporated to facilitate creativity in a co-creative system.
We illustrate the framework components with an AI assistant tool for package design, and a case study of perfume bottle design (leveraging a created dataset of \textasciitilde{}24K bottle images).  Currently there is a lack of AI tools for package design, despite the huge impact package design has on product sales \cite{OMMarketing2008,pepsico,nielsen,REIMANN2010431}.  Although our  tool uses images as designs, the ML models are generally applicable to other domains.  Further, under the idea of diagrammatic reasoning \cite{cheng2001cognitive}, \textit{i.e.}, reasoning via visual representations, creativity systems that operate on images can potentially be applied to non-visual domains \cite{liapis2016can}. 

\begin{figure}[h]
\centering
\includegraphics[width=0.95\textwidth]{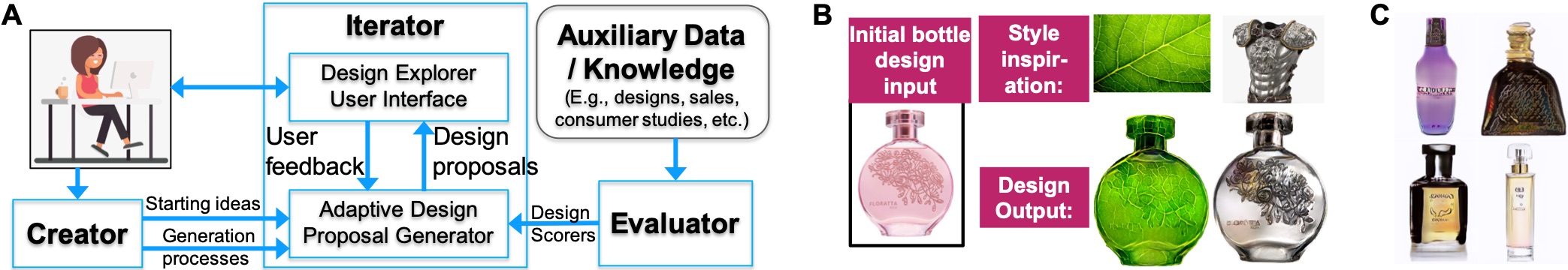}
\caption{(A) Framework overview (B-C) Examples of design modification and generation, resp., in case study}
\label{fig:combo1}
\end{figure}


Our framework (Figure \ref{fig:combo1}A) consists of 3 key components, leveraging ML with collected design data.  \emph{\textbf{Creator}} comprises a set of ML tools for generating designs and variants.  Generative modeling, \textit{e.g.}, generative adversarial networks (GANs) \cite{goodfellow2014generative}, is used to capture the constraints or manifold of the design space, and generate realistic, novel designs for exploration / design proposal, with further constraint through filtering or score-optimization.  Our tool includes this - Figure \ref{fig:combo1}C shows random, novel designs generated by a progressive GAN \cite{karras2017progressive} trained on our data.  Additionally, conditional versions \cite{mirza2014conditional} can be used for controlled generation.  Generation can be conditioned on category or user-selected design features, or parts of the design can be fixed, using image-completion style GANs \cite{Yu_2018_CVPR}.  Sketch-to-image (\textit{e.g.}, \cite{isola2017image}) can be used to both allow rapid prototyping and refinement with a designer-provided sketch, or to fix the design structure (\textit{e.g.}, using the edge map) and fill in different color palettes (part of our tool).  Similarly design structure can be fixed while allowing modification of style (\textit{e.g.} color and texture) through style transfer \cite{gatys2016image,jing2019neural,Huang_2017_ICCV} - implemented in our tool (Figure \ref{fig:combo1}B).  Furthermore, creativity can be infused in generative models to expand the search space \cite{elgammal2017can,Sbai_2018_ECCV_Workshops,das2019toward}.

\emph{\textbf{Evaluator}} leverages data to predict design modalities like novelty (\textit{e.g.}, \cite{ding2014experimental,wang2018generative,hendrycks2016baseline}), cost, and consumer appeal, leveraging modern, highly-accurate predictive modeling \cite{chollet2017xception} and transfer learning (\textit{e.g.}, \cite{donahue2014decaf,sharif2014cnn}) to prime modeling off the existing design set.  Each evaluator provides one or more values or \textit{scores} that characterize a design, and it is up to the user to decide its influence.  In our case study we only include aesthetic modalities of shape and color. \emph{\textbf{Iterator}} works with a human user to evolve design concepts and explore the design space, iteratively proposing candidate designs (including using Creator components) and learning user preference based on which designs are selected (and adapting future proposals) - leveraging again transfer learning to speed up the learning process, and also reinforcement learning \cite{sutton2018reinforcement} techniques.  Evaluator modality scores feed into the user preference model as well and also help guide the user's feedback to the Iterator. 

\begin{figure}[h]
\vspace{-2mm}
\centering
\includegraphics[width=0.99\textwidth]{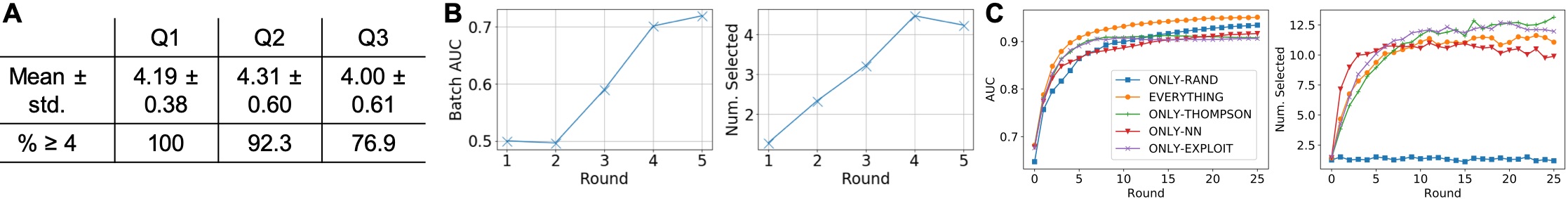}
\caption{(A) User response stats (B-C) AUC and num. selected per round for human and artificial users, resp.}
\label{fig:expres}
\end{figure}

We implemented a version of the framework as a creative assistant tool for perfume bottle design.  We used a simplified Iterator interface that displays a 3x6 grid of 18 design proposals each round.  Users select designs they like, clicking ``SUBMIT'' when finished, then the Iterator updates its model of learned user preference and produces a new set of design suggestions.  We tested different strategies for design proposal, balancing between exploring the design space to inspire the user and selecting results in-line with the current, learned design goal.  We evaluated the Iterator quantitatively first by plugging in an artificial user with fixed, deterministic preference (\textit{e.g.}, for thin bottles as measured by aspect ratio).  We measured the accuracy (in terms of Area Under the Curve - AUC) of the learned preference model on a held-out test set as well as number of designs selected per round, for the different proposal methods.  Results for the ``thin'' concept are shown in Figure \ref{fig:expres}C. We found the system was able to efficiently learn these simple fixed concepts, with a combination of proposal strategies providing efficient learning while still enabling exploration.  A purely random proposal strategy (``ONLY-RAND''), however, led to a much lower rate of user-selection, which could frustrate a human user, and slightly lower learning rate.  

We also performed a user study with 13 non-expert participants, assigning each a random bottle design task defined by a perfume description (such as ``the smell is sweet, fruity, girly, and flirty'') - as is common in real design specifications.  We demoed the design-assistive components of the tool for each participant, and had each try the Iterator for his/her task (with a combination proposal strategy).  We recorded mean proposal set accuracy (batch AUC) of the learned preference model and number of proposal designs ``liked'' per round.  We found even for this more realistic, complex case the preference was able to be learned fairly accurately and the tool quickly increased the number of proposal the users liked (Figure \ref{fig:expres}B).  We also asked 5 questions about their experiences.  On a scale of 1 to 5 (5 being the most): \textbf{(Q1)} ``How useful do you think the whole suite of tools would be to help in your design process for coming up with a new, creative design?''; \textbf{(Q2)} ``How much did it help you discover candidate designs related to the task?''; and \textbf{(Q3)} ``How much did it help you explore the wide space of designs and stimulate your creativity?''; (free response) \textbf{(Q4)}: ``What did you like most about the tool?'' and \textbf{(Q5)}: ``What would you most want to improve about the tool?''.  

Summary statistics of their responses to Q1-3 are given in Figure \ref{fig:expres}A.  We received largely positive responses, users seeming excited about the capabilities.  Users felt the sketch-to-design and style transfer components could be very helpful and time-saving.  For the Iterator, users liked its ease-of-use and felt the exploration aspect was useful, providing unexpected ideas at times.  However, users felt there were sometimes too many similar designs presented - a result of modeling the proposals per-image (set-based metrics like diversity will be added in future work).  Having more capabilities to control the exploration was a key desire - such as the ability to fix the color theme, or take aspects of designs they liked, like shape or color only.  We plan to add such capabilities as part of future work, by incorporating more Creator components mentioned, such as conditional models, in our tool.

\section*{Ethical Implications}
One might think a tool capable of generating creative designs could take work away from designers.  However, the key point here is that it is a \emph{co-creative} assistant - it does not replace designers, but instead works with designers, making their jobs easier and more enjoyable, and stimulating their creativity to new heights.  As such it will make designers more efficient, spending less time on trivial tasks and spending more on the creative ones, making it easier to consider other data (such as cost, sales impact, or competitor data) in their designs, and helping them handle more design tasks without fatiguing their creativity.  Another point is it facilitates cross-company participation in design, helping others in a company, like marketers and production teams, have a more productive collaboration with designers and possibly contribute in the design process.  

Since the tool is ML-based and requires existing designs to train the model, the use of the design data could be considered an issue, if such data from one company is used to train models used in a tool offered to competitors.  In some domains, many designs are public (such as publicly sold perfumes), but each company may have even more non-public designs with various data on them, such as consumer studies.   However, one ethical way of providing such a beneficial tool to all designers is giving companies the choice to opt-in to a shared-data model.  I.e., they could choose to keep their designs internal - using only their own and public designs for model training, and not leveraging competitor designs as well.  Or they could instead choose to share some set of their internal designs to aid in general model training, and also benefit from designs supplied from other competitors.

\bibliographystyle{IEEEtran}
{\small
\bibliography{reference}}

\clearpage
\section{Supplementary Material}

\subsection{Co-creative assistant for packaging design use-case motivation}
Packaging (e.g., bag, box, or bottle) design is an integral part of product and brand identity, which influences consumer experience, buying decisions and loyalty \cite{OMMarketing2008}.  In the age of differentiating consumer experiences, good packaging designs strive to offer unique appeal in terms of aesthetics, utility, creativity, trendiness and brand impression while keeping costs and environmental impacts at bay. There are ample examples that indicate the impact of package design on consumer impressions. For instance, according to \cite{nielsen}, the right packaging may boost sales for an average brand by 5.5\%.  \cite{REIMANN2010431} found aesthetic packaging can boost sales for a new brand, and \cite{Rizwan2014} found the preferences for packaging colors are very different between the West and Far East. 
The consumer industry often revamps packaging for common-use products frequently to inject new life into sales. For instance, Pepsico achieved sales growth of 35\% to 50\% in some countries for its Gatorade brand with new structural bottle designs \cite{pepsico}.

Despite packaging's critical role, current packaging design processes lack data- and AI-driven tools to facilitate and enhance these processes.  Based on internal discussion with multiple consumer industries, in current design work flows for packages, marketers, product developers and designers often work in silos, wasting time in back-and-forth communication and iterative manual design refinements.  Design evaluations are performed on multiple real trial prototypes with consumer testing making the entire process excessively expensive and lengthy. There is no data-driven design evaluator that allows users to estimate how a design would perform along different dimensions, such as sales, appeal to customer segments, cost, etc.  Furthermore, the design process itself could benefit from assistive tools.  Currently a non-designer with some ideas has to rely on a designer to create visualizations and there is no tool that allows the non-designer to make quick design tweaks and visualize her ideas.  Even an expert designer may get trapped in past design biases or blind spots, and may miss potential blockbuster new designs. 

We developed our ML-based co-creative framework, in part motivated by these challenges, and specifically chose the package design use-case as our target one to demonstrate the framework.  The framework, and our realization of it as an AI-assistant tool, assists human designers to rapidly produce more effective and creative package designs without taking creative control away from the designer. Akin to these implicit phases in a typical design flow, our framework consists of three broad sets of tools, namely Creator, Evaluator and Iterator. Creator can take some initial inspiration seeds from a human designer (e.g., color palette, shape, inspiration images, etc.), enables rapid design space exploration by learning the latent design manifolds, and generates new designs. Evaluator enables evaluating generated designs along different dimensions, extracting different modalities of designs in the form of multi-dimensional metrics, e.g., learning relationships between design aspects and outcomes, such as aesthetics based on shape and color or novelty compared to competitors. It is able to ingest or leverage additional data or knowledge, such as consumer studies from marketing, cost constraints from product development divisions, and knowledge about shape learned from large data collections. Finally, Iterator uses an interactive UI to propose candidate designs along with their metrics to a user and gather feedback on the designer's current preferred choices. Iterator further learns a model of the designer's preferences for her current design goal and guides the search in the design space, facilitating the next iteration of creation and evaluation.   The framework and tool are flexible in that human users with different skill levels and needs can use parts of it for different purposes. For instance, an expert designer may use Evaluator to estimate her design's appeal and Creator or Iterator to get inspiration for out-of-box designs, while, a marketing professional may use the whole framework to tweak an existing design and visualize the results. 

We demonstrate the feasibility and potential of this framework with a use case on designing perfume bottles. We show that our framework enables facilitating and enhancing the design process and grounding it in real data so both designers and non-designers can explore design possibilities and modifications for designs as desired and get quick mock-ups without long delays. Besides package design, we believe that our general framework is also applicable to broader design applications such as logo and website creation, advertisement, and others, as mentioned in the main paper.  Even non-visual domains can use the same underlying machine learning models and framework components and structure, and diagrammatic reasoning is also an option.
\subsection{Model Implementation for Package Design}
In this section we will go over the framework again and specifically how it applies in the context of package design (and the implementation we used for our tool to demonstrate the framework).  We reproduce some figures with larger size and provide more details from the main paper.

\subsubsection{Overview}

\begin{figure}[h]
\centering
\includegraphics[width=0.75\textwidth]{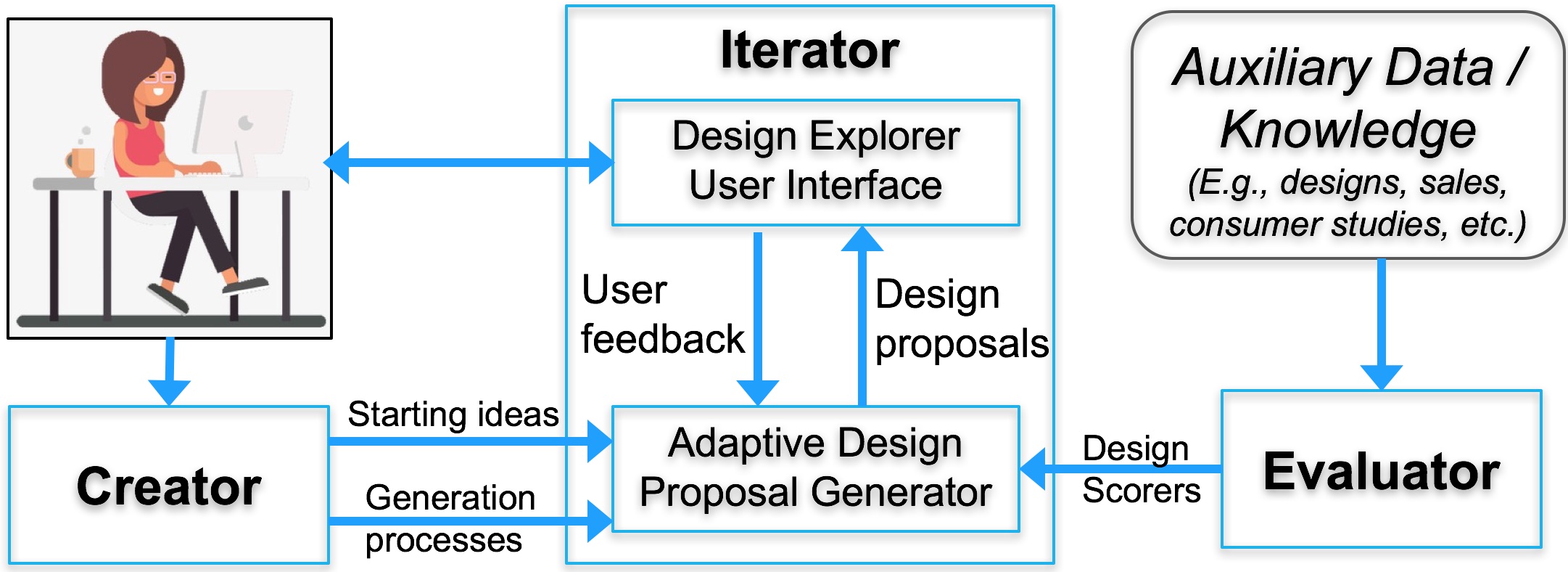}
\caption{Overview of proposed interactive AI design framework.}
\label{fig:overview}
\end{figure}

Our human-AI interactive design framework consists of three key components as illustrated in Figure \ref{fig:overview}.  Creator collectively comprises a set of tools to enable the user to create design realizations from more primitive components as well as refine or modify designs, and also includes processes for generating designs or design variants.  Evaluator consists of scorers that can characterize a design along different dimensions (like shape, color, cost, customer appeal, etc.). The outputs from Creator and Evaluator feed into Iterator - which enables the user to interactively explore the design space, by suggesting relevant candidate designs meeting her preference and current design goals along the multiple modalities, and iteratively improving design suggestions and the learned model of user preference.  The end result of the system is to quickly provide the user with design inspirations and rough realizations that can be used to base final designs on.  

\subsection{Data}
\label{sec:data}
To show results for each component of our framework, and to evaluate the interactive system, we created a dataset of 24,126 perfume bottle product images (see Figure \ref{fig:round5} for a sample).
We first collected around 35,000 product images from the internet and filtered this to clean, bottle-only images using a classifier we trained on a small set of manually labeled images, leveraging transfer learning with a pre-trained image classification network \cite{donahue2014decaf,sharif2014cnn}.  

We also generated masks of the bottles to separate bottles from background by performing Canny edge detection, followed by Gaussian filter blurring and flood filling the image up to the blurred edge boundaries.  This is used to determine characteristics of images like dimensions, and mask results after image operations like style transfer.
\subsubsection{Creator}
\label{sec:creator}

In many domains, a typical design process begins with a design brief which is a document that outlines the deliverables and information of the project including background, marketing message, target audience etc. It provides a designer with inspirations, directions, and constraints for the design. 
For example, in packaging design, a brief can indicate a primary color for a package, \textit{e.g.}, ``The handsome deep blue was chosen because it conveys a sense of tradition and strength, which is right for XYZ'' (XYZ is a cologne for men). 

Creator consists of several tools to suit different scenarios in this design phase. It can start from some initial inspiration seed (e.g., color palette or shape), explore through the design space rapidly by leveraging learned latent design manifolds, and generate new designs. 

\textbf{Rapid concept exploration with sketches:}  
Many designers start with sketches to capture their initial ideas and spend a significant amount of time to refine, detail, and explore variants of those designs. It is time-consuming but often the only way for a designer to determine if concepts are worth exploring further. One tool in Creator assists the human designer to explore through the design space rapidly with sketches.  The tool takes sketches and/or color palettes as input, and renders realistic designs. While those designs are still semi-finished, they give the designer the look and feel of different concepts, and enable him to focus more on rapid formulation of ideas. 

\begin{figure}[ht]
	\centering
	\begin{subfigure}[t]{0.7\textwidth} 
		\includegraphics[width=\textwidth]{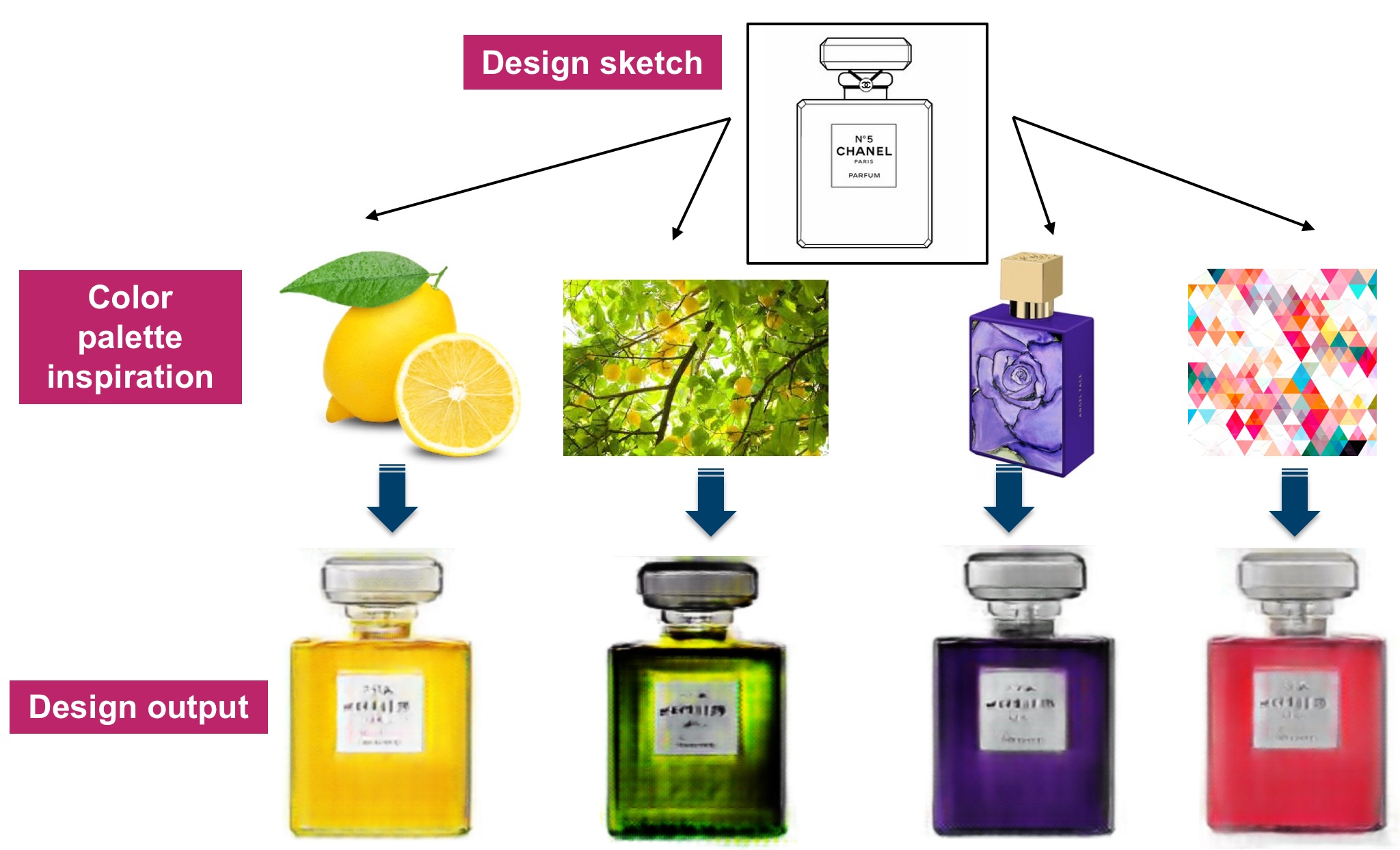}
		\caption{} 
	\end{subfigure}
	\hspace{1em} 
	\begin{subfigure}[t]{0.2\textwidth} 
		\includegraphics[width=\textwidth]{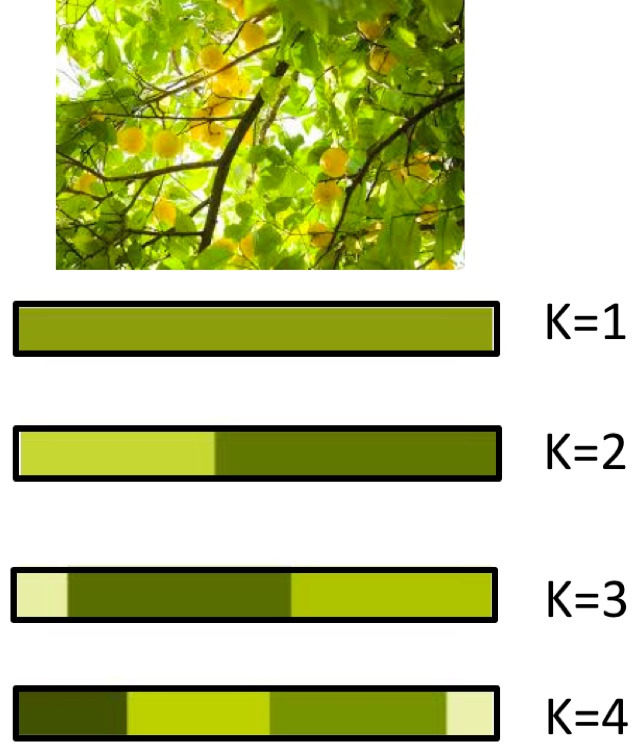}
		\caption{}
	\end{subfigure}
	\caption{{(a)} An example of Creator on perfume bottle designs based on inspiration images. {(b)}: Color palettes generated from an inspiration image for different number of clusters.  These feed into the conditional GAN to guide the bottle sketch completion.}
	\label{fig:sketch}
\end{figure}

The framework behind this tool is adapted from Pix2Pix \cite{isola2017image}, which uses a conditional generative adversarial network (cGAN) to learn a mapping from an input image (corresponding to the image edge map) to an output image (corresponding to the complete image). 
Differentiating our approach, we make a key modification to the model for our application: our tool takes additional images as input from which a color palette is extracted and fed into the generator to influence the color in the generated result (see Fig \ref{fig:sketch}(a) for an example of some results).  This additional input is crucial for package design as clients often specify a color palette or provide color inspirations in the design brief.  In addition this approach could be used to enable constrained generation for exploration while fixing the structure, but exploring different sample colored palettes - and could also be extended to have multiple color palettes corresponding to different parts of the design.


To obtain the color palette, we first run K-means to cluster the pixel intensities of an RGB image, then create a histogram which depicts the dominant colors represented by each centroid, where the width for each color is proportional to the number of pixels in that cluster. Fig \ref{fig:sketch}(b) shows various color palettes generated from the same image with respect to different cluster sizes.  Each centroid represents a dominant color in the inspiration image.  Pixels that belong to a given cluster will be more similar in color than pixels belonging to a separate cluster, and the number of dominant colors is determined by the cluster size which is an input parameter of K-means.  The color palette of each training bottle image is fed into the conditional GAN model as part of the conditional input, along with the edge map of the image, during training, and when applying the model with a bottle sketch the extracted color palette from inspiration images are used.

\begin{figure}[h]
\centering
\includegraphics[scale = 0.18]{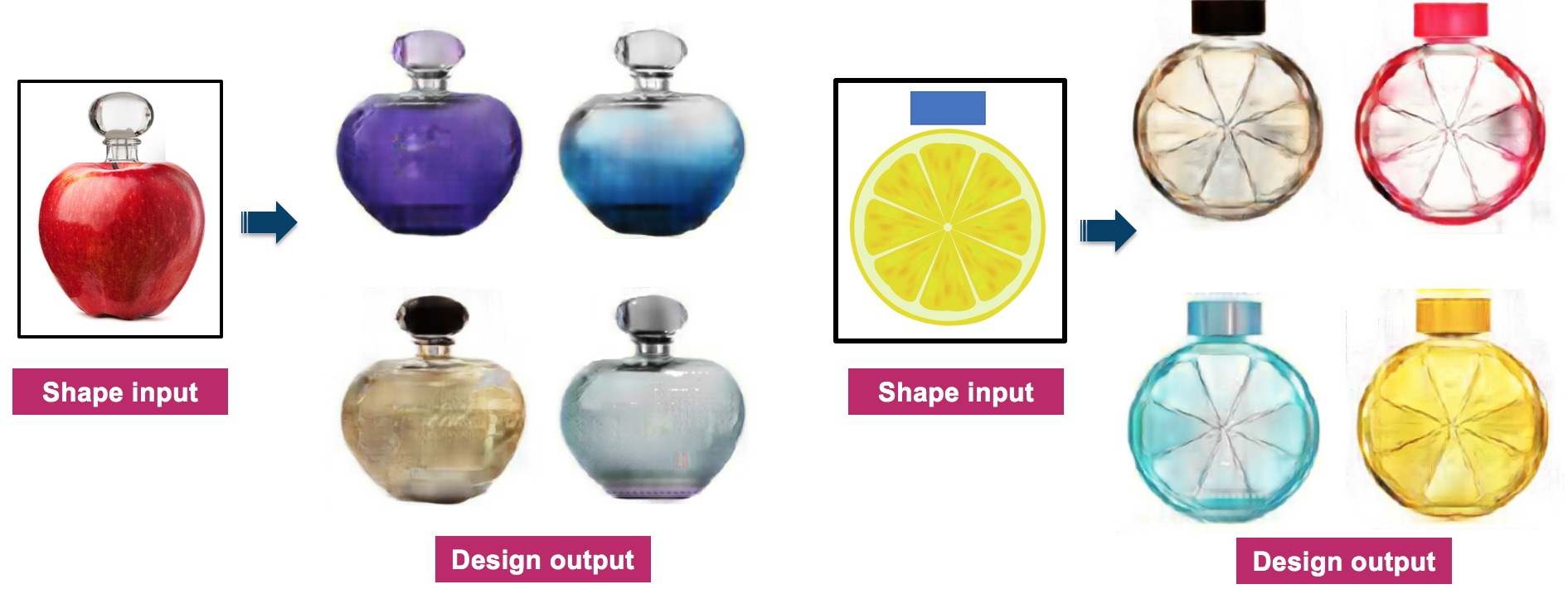}
\caption{Examples of Creator with abstract shape input.}
\label{fig:pix2pix2}
\end{figure}

Figure \ref{fig:pix2pix2} shows two more examples to showcase the robustness of this creation tool. Instead of a sketch of a bottle, the ``inspirational shapes'' are given by an assembly of an apple and a glass cap in the first example, and a lemon slice clipart and a rectangle in  the second. These examples demonstrate that even without a formal sketch, the tool is capable of producing realistic-looking designs from quick and rough shape input given by a designer.  

\begin{figure}[h]
 \centering
 \includegraphics[scale = 0.1]{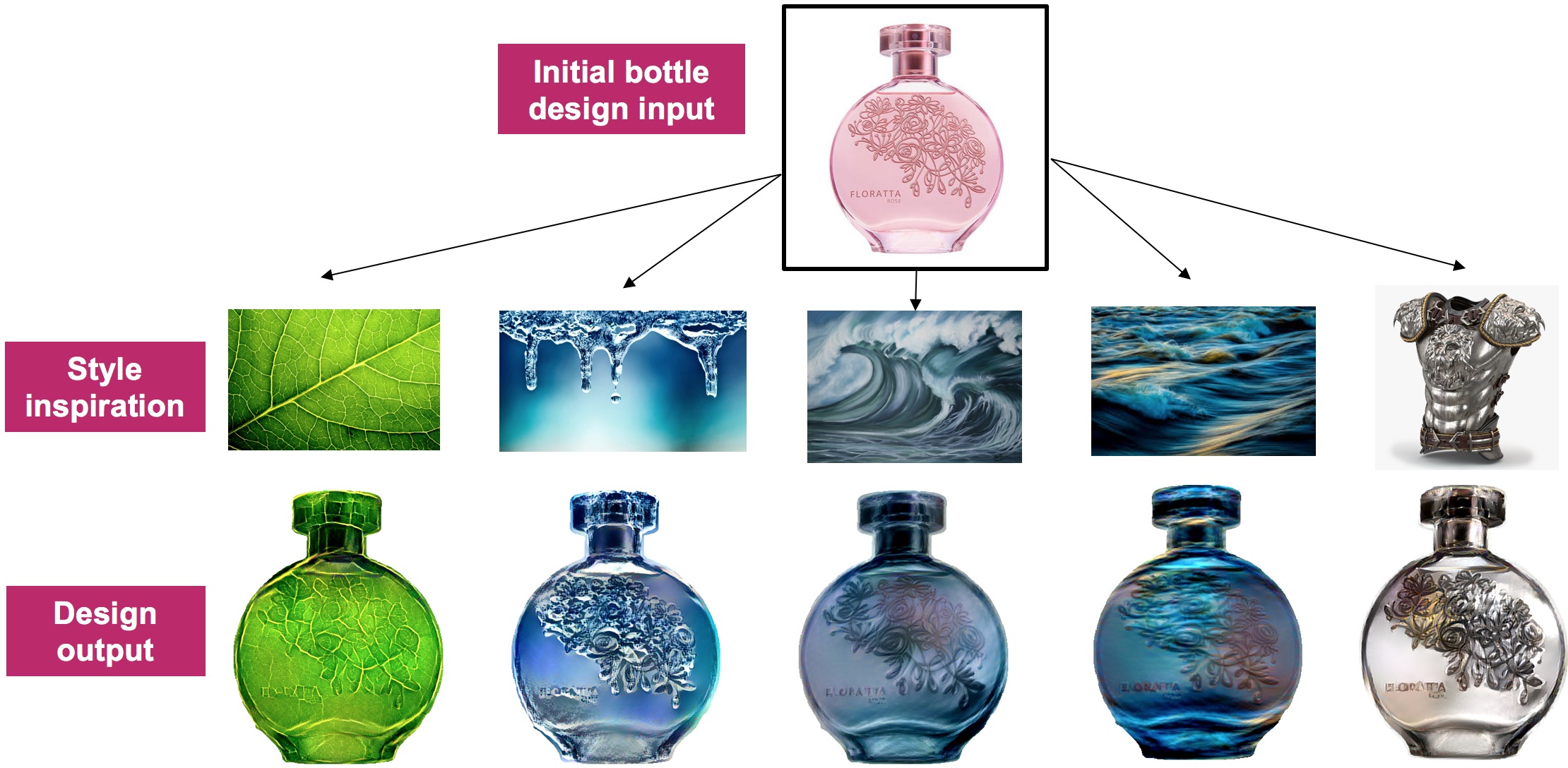}
 \caption{Examples of flanker bottle designs with style transfer.}
 \label{fig:styletransfer}
 \end{figure}

\begin{figure}[h]
 \centering
 \includegraphics[width=0.9\textwidth]{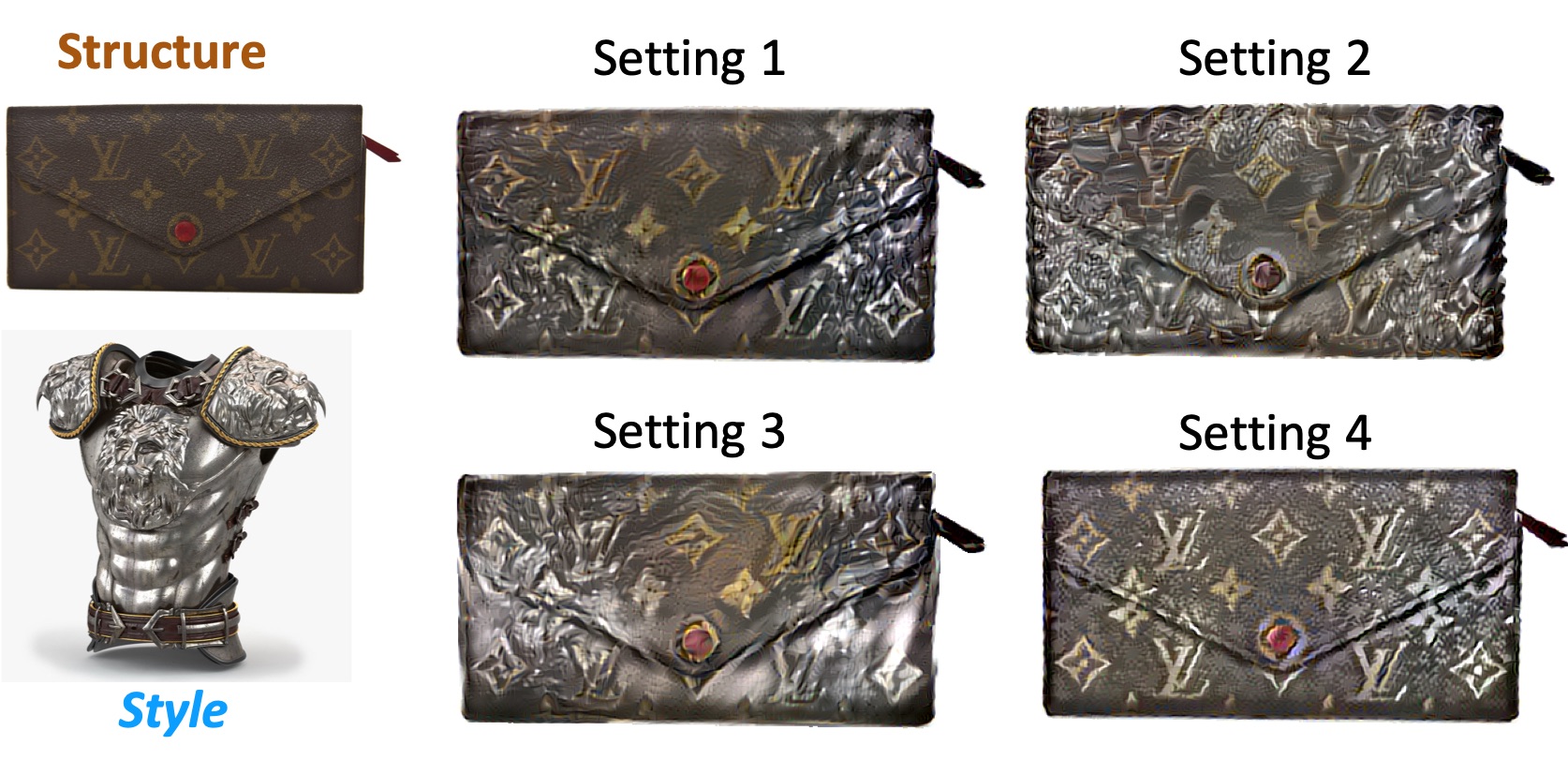}
 \caption{Examples of flanker bottle designs with style transfer for different packaging type: wallets.  Each result is generated with different settings for the style transfer, generating how style transfer can contribute to expanding the search space of designs, even without vaying style images.}
 \label{fig:flanker2}
 \end{figure}

\textbf{Style transfer and flanker design:}
Borrowing a perfumery concept, a flanker refers to a newly created perfume that shares some attributes (e.g., packaging or notes) of an already existing perfume. 
These attributes may be the name, packaging or notes of the existing fragrance. 
For example, Dior's 1985 fragrance Poison was followed by Tendre Poison (1994), Hypnotic Poison (1998), Pure Poison (2004), Midnight Poison (2007) and Poison Girl (2016), all with roughyl the same packaging in different syles. 
\begin{figure}[h]
\centering
\includegraphics[scale = 0.25]{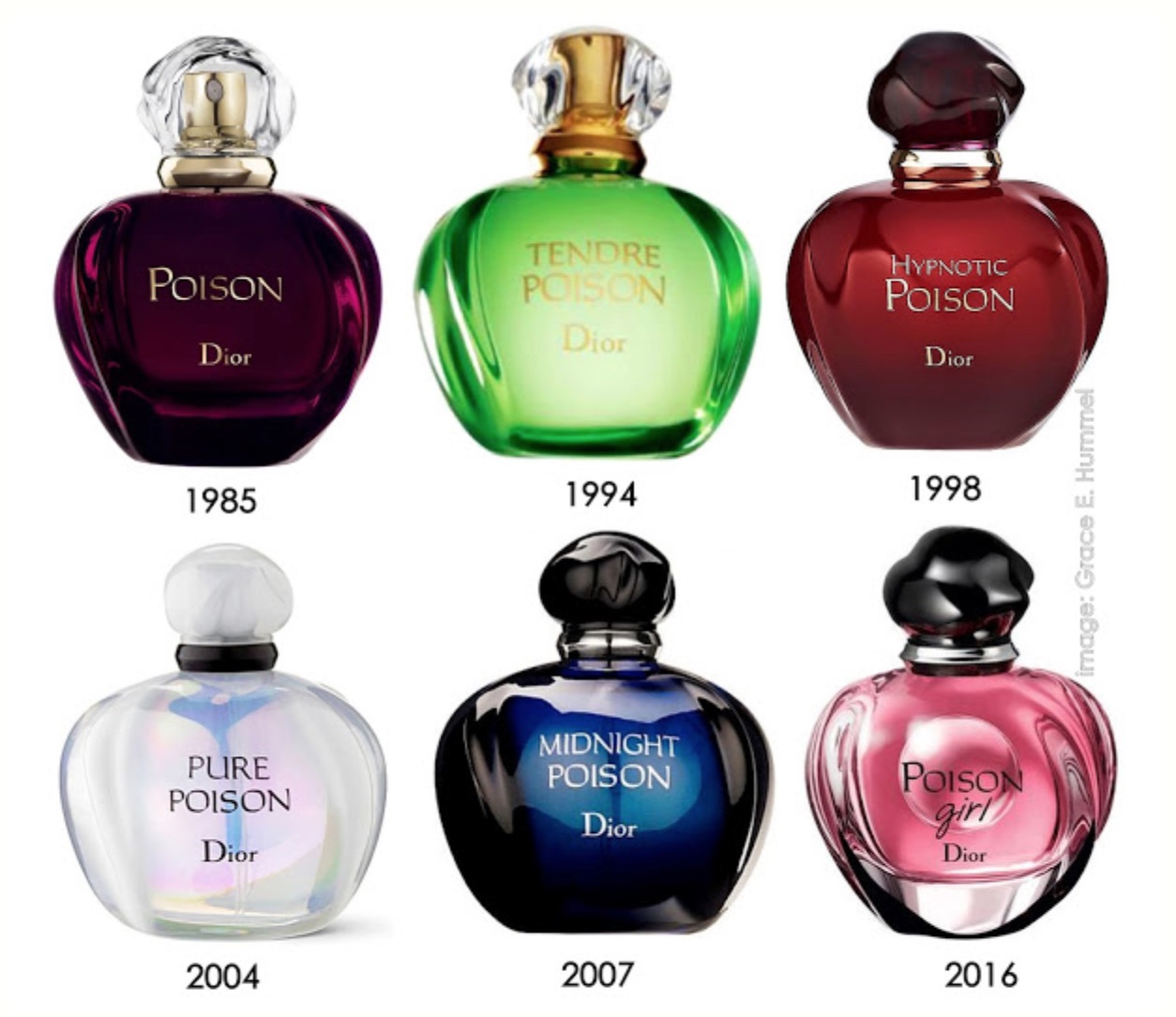}
\caption{Flankers for Dior Poison. Image source: $https://cleopatrasboudoir.blogspot.com/2013/05/poison-by-christian-dior-c1985.html$(We can include this image in the supplementary material)}
\end{figure}
In terms of flankers' packaging, they tend to share very similar bottle designs, but differ in terms of colors, texture and potentially material
. For this task, we utilize the approach of neural style transfer \cite{gatys2016image} to copy the style from a style image and apply it to a content image representing the design we want to modify the style of. 
When it is applied to flanker package design, the content image can be the an existing design (e.g., perfume bottle), while the style could be an inspiration image from the design brief. 
Fig \ref{fig:styletransfer} depicts examples of style transfer with our data.  For example, the last column is an example of a flanker design where we transfer the style of shiny armor to a perfume bottle, creating a ``masculine'' version of the existing bottle.  Style transfer can also be used to generate design variants to help explore the design space and provide creative design proposals to the human user in the co-creative system.  The results of style transfer can even be varied based on changing the settings (as illustrated in Figure \ref{fig:flanker2}), adding another dimension of generated design variation.

\begin{figure}[h]
 \centering
 \includegraphics[width=0.7\textwidth]{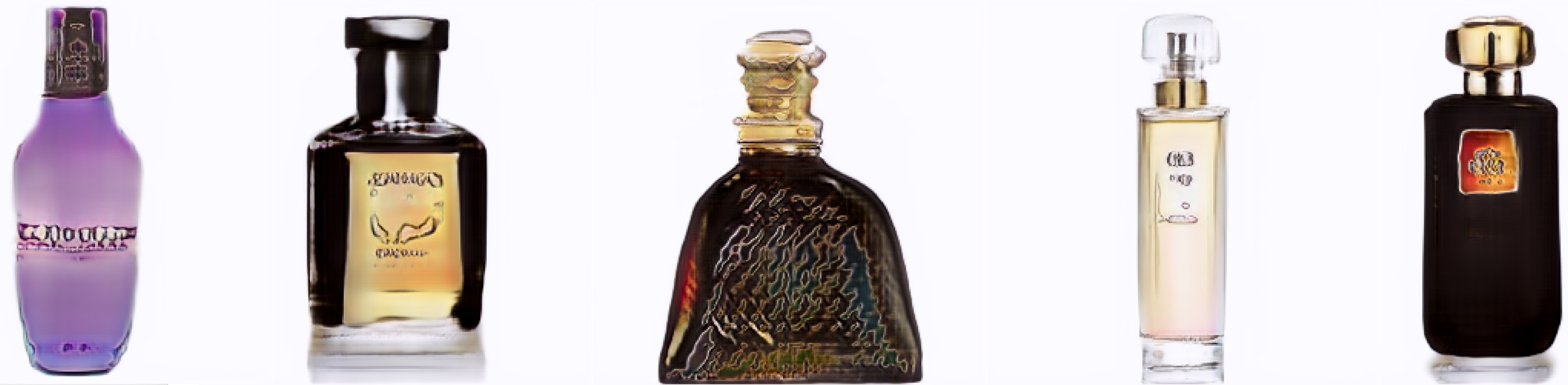}
 \caption{Examples of random, new designs generated.}
 \label{fig:pgan}
 \end{figure}

\textbf{Design manifold exploration with generative modeling:}
\label{sec:design-manifold}
The space of design specifications, such as images depicting a design, is huge and  difficult to explore directly.  However, designs exist on some lower dimension manifold encompassing the inherent design characteristics and constraints, such as symmetry, specific cap and body shapes and locations, etc.  By learning to model and generate designs in this low-dimension space we can facilitate design space exploration, e.g., to find similar designs, interpolate between designs, and even generate completely new designs.  Our system uses generative neural network models to accomplish this.  For example, we trained a progressively grown GAN (PGAN) \cite{karras2017progressive} on our collection of bottle images.  An example of 5 random, non-cherry-picked designs generated by this trained network are shown in Figure \ref{fig:pgan}.  We found it is able to generate fairly realistic, novel designs, quite dissimilar from ``neighboring'' training-data designs in the embedding spaces.   Along with the other Creator components, and conditional GANs, this can be used as a key component for design space exploration described in Section \ref{sec:iterator} - for generating new and nearby designs to help flesh out and fully explore the design space.  

\subsubsection{Evaluator}
\label{sec:evaluator}
Evaluator consists of a number of scorer systems each evaluating the design along a given dimension, or \emph{modality}, trained using relevant data for that modality.  For example, a scorer may be a predictor estimating the cost of a design based on design specifications and associated manufacturing cost details, or predicting the appeal to different demographics based on consumer studies and sales data, or quantifying novelty from an existing database of designs.  
Each scorer takes a design specification (e.g., an image representing the bottle design), and maps it to one or more values representing the design along its modality, e.g., a single estimated cost value, or a customer preference profile vector, or a shape representation vector, etc.

For our proof of concept we focused on two key, universal design modalities - shape and color.  To represent the shape modality, we use high-layer outputs of a pre-trained image classification neural net trained on a large image data set (ImageNet) as these have been shown to provide useful general image features \cite{donahue2014decaf,sharif2014cnn}, and style transfer work suggests high-level layer outputs capture image structure (i.e., shape) well \cite{gatys2016image}.  Specifically we use the pre-trained Xception network's global pool output layer \cite{chollet2017xception} - a 2048-D vector. 

To represent the color modality, we use a simple a 4-D descriptor, which we found to be effective at capturing dominant color themes.  We transform each pixel to the HSV space \cite{joblove1978color} and convert the radial hue value to its 2-D coordinate on the unit circle.  We then take the average of these pixel vectors across the image, resulting in a simple 4-D descriptor.

\subsubsection{Iterator}
\label{sec:iterator}
The Iterator component enables interactive exploration of the design space by iteratively proposing a set of candidate designs to a user, collecting feedback, and updating a learned model of the user's current preference to refine the proposals for the next round.  It leverages a set of different proposal methods, trading off between exploring in different ways and exploiting current learned knowledge of the user's preference.  Exploration can help inspire and lead the user to creative designs, while exploitation is necessary to rapidly propose relevant results.  The overall process is given in Algorithm \ref{alg:iterator}.
\begin{algorithm}
\caption{Interactive Design Exploration}
   \SetKwInOut{Input}{input}\SetKwInOut{Output}{output}
   \Input{\# of candidates per proposal method}
   \Output{Set of selected candidate designs}
    Generate random set of candidate designs, $\mathcal{C}$\;
    \While{User does not terminate}{
        Show user set of candidate designs $\mathcal{C}$\;
        Retrieve feedback from user \;
        Update user preference models from feedback\;
        (Optional) update generator to align with user preference\;
        Empty candidate design set $\mathcal{C}$\;
        \For{each proposal method}{
           Generate proposals from method and add to $\mathcal{C}$\;
        }
    }
 \label{alg:iterator}
\end{algorithm}

Iterator consists of 5 components.  First is an interface to present proposed designs and obtain user feedback.  In our proof of concept, the feedback is just a simple ``like'' or not and the interface is a graphical user interface visually displaying candidate designs (Figure \ref{fig:round5}).  In general the interface can provide more details per design and the feedback can be more nuanced, including ratings along different modalities.  The second component is a generator, for generating candidate designs.  This can be on-the-fly generation using the Creator (e.g., generative models, style transfer, etc., with random inputs).  Alternatively, candidate designs can be generated and collected in advance and simply sampled from (i.e., a pre-built database of designs).  The generation can also be biased using the user preference model or design starting point inputs. The third component is the set of Evaluators (see Section \ref{sec:evaluator}) - providing embeddings of the design along different modalities.  These encode key design dimensions users may care about, so the user preference models are based off of them, and they can also be incorporated in the UI to inform and guide the users. 

The fourth component is a trainable user preference model, i.e., a predictive model of the user's design preference, in the current design session.  For the case of ``liking'' a design or not, this is simply a probabilistic classification model. This model is updated after each round of proposals and feedback from the user, via incremental learning \cite{losing2018incremental}, or simply re-training, which is typically quick enough due to the small sample size.   This model can be based on both the full-fledged design representations (\textit{e.g.}, the design images in our package design use case) and the Evaluator modality outputs, or for simplicity it could be restricted to be based off of only the modality outputs (although this could limit its potential accuracy and ability to account for modalities not captured in the given set).

Finally, the fifth component is a set of proposal methods.  These involve sampling from or searching over the design space (which can be represented by the Evaluator modality embeddings for simplicity and efficiency), and vary over providing design space exploration, enabling rapid user-preference learning, and exploiting learned user preference.  The following proposal methods are used. 

\noindent\textbf{RAND:} This method randomly samples from the generator inputs/database of designs.  We also implemented a variant that samples from the distribution imposed by the user preference model via rejection sampling \cite{robert2013monte}.

\noindent\textbf{EXPLOIT:} A larger set of designs (e.g., a few hundred) are sampled via RAND and the top scoring designs according to the user preference model are taken for the proposals.

\noindent\textbf{THOMPSON:} Based on a reinforcement learning perspective \cite{sutton2018reinforcement}, for each candidate proposal it samples a user preference model from an estimated distribution over preference models and takes the top-scoring design from a RAND sample.  To capture the distribution over models, bootstrap sampling is used (modeled after the approach in \cite{riquelme2018deep}). A set of $k$ different models are maintained and each has probability $p$ of being updated with a labeled design after each round - so that each model is incrementally trained on a different, random subsample of data. 

\noindent\textbf{NN:} This proposal method samples designs in nearby neighborhoods of selected designs in the previous round - in different modality embeddings.  This enables quickly seeing similar designs and design variations to ones the user liked, to inspire the user and rapidly expand the set of relevant designs proposed.  
\subsection{Experiment}
\begin{figure}[ht!!]
	\centering
 		\includegraphics[width=0.75\textwidth]{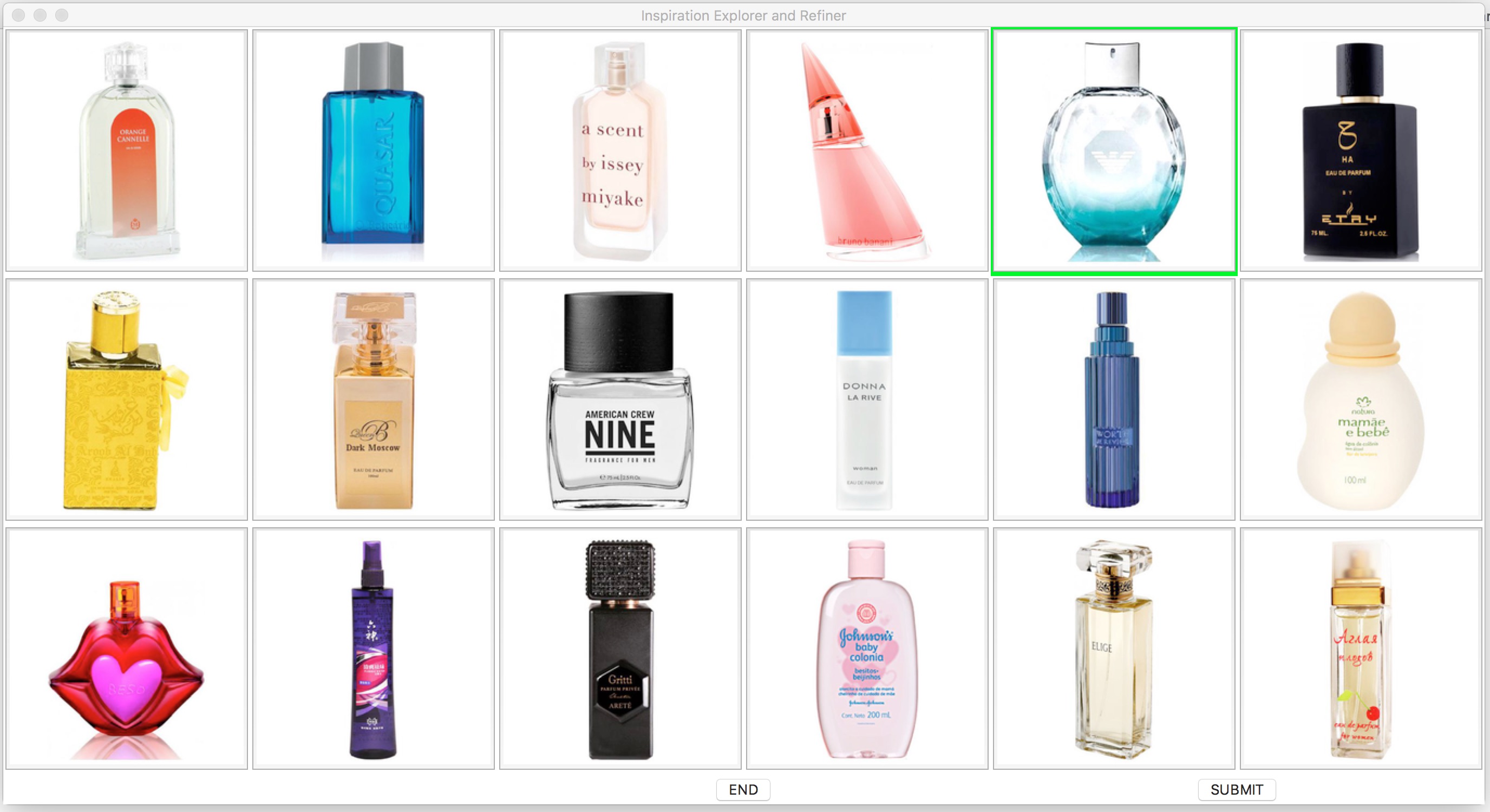}
 		\caption{Starting sample of designs } 
 		\label{fig:design-sample}
 		\end{figure}

\begin{figure}\centering
		\includegraphics[width=0.75\textwidth]{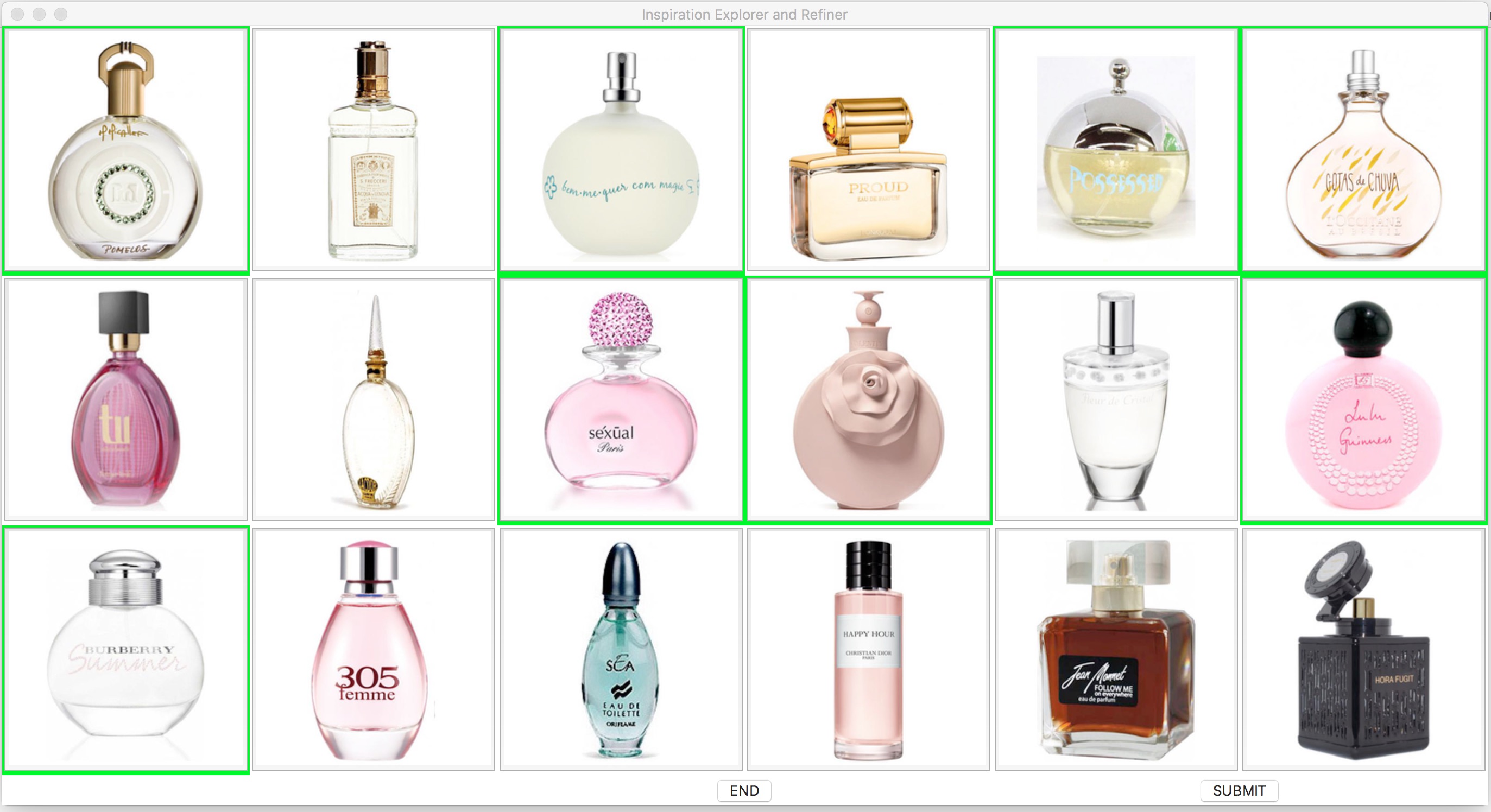}
		\caption{User interface - at the 5th round, selections in green.} 
		\label{fig:round5}
\end{figure}

We focus on most-thoroughly evaluating the Iterator component of our proposed framework, as it includes the other components and lends itself to quantitative and qualitative study.  We implemented and tested the Iterator approach for simple binary selection feedback.  We focus on the case of perfume bottle designs as represented by bottle product images (data described in Section \ref{sec:data}).  Screen shots of the interface are shown in Figures \ref{fig:design-sample} and \ref{fig:round5}.  When a user clicks to select a design, it is highlighted in green.  After finishing selections in the current round, the user clicks the ``SUBMIT'' button to submit the selections, and clicks the ``END'' button when finished exploring designs.  We focus on the representative approach of using a pre-generated database of designs, as described in Section \ref{sec:iterator}, for simplicity.

\subsubsection{Setup}
We test the ability of the system to aid in discovering and learning a number of design concepts.  We fix each design concept in advance then interact with the AI system through the interface to see how quickly it can learn the concept and propose relevant designs.  For each experiment run we plug in a program to simulate human operation according to fixed scoring rules for selecting designs.  

\textbf{Design concept tasks:}
In order to avoid overly subjective results, we use 4 simple design concepts, scored automatically by fixed rules.  The first two tasks are \textbf{RED} and \textbf{BLUE} - which represent the concept of dominant red or blue colored bottles, respectively.  These are scored by converting each RGB pixel value to the difference between the red / blue channel and the maximum value of the other channels, and averaging over all pixels.  If this score is then greater than an outlier threshold, it is classified as a positive case.  Similarly the next two concepts, \textbf{FAT} and \textbf{THIN}, are also based on a threshold rule - in this case on the aspect ratios (width divided by height) of the bottles, which are determined by extracting the width and height of each bottle using masks derived as explained in Section \ref{sec:data}.  THIN corresponds to very tall, skinny bottles, and FAT to very short, fat bottles.

Additionally, to better simulate human interaction of sometimes selecting cases that seem close to the target concept - for each concept we set another, smaller threshold that includes roughly 2000 additional bottles.  We then assign a probability using a sigmoid function of the bottle's score normalized between the two thresholds, and randomly label these bottles according to the probabilities in each run.  
The final percentages of positively labeled cases are shown for each automated task in Table \ref{tab:num-pos} - where ``Always'' indicates the percentage of bottles that are always labeled positive, according to the first, tighter threshold, and ``Average'' the average number labeled positive across all runs - including the randomly labeled ones.  We also illustrate the results of manually using the tool for a fifth design concept of ``circle-shaped bodies''. 
\begin{table}[htb]
\scriptsize
\centering
\begin{tabular}{c|c|c|c|c}
\textbf{Task} & RED & BLUE & FAT & THIN \\\hline
\textbf{Average} & 9.2\% & 2.1\% & 9.5\% & 7.5\% \\\hline
\textbf{Always} & 5.5\% & 0.5\% & 7\% & 3.6\%
\end{tabular}
\caption{Percentage of positive-labeled cases per automated task.}
\label{tab:num-pos}
\vspace{-4mm}
\end{table}

\textbf{System setup and procedure:}

Our interface provides 18 proposals to the user each round in a 3x6 grid.  We compare 5 different proposal methods.  The first 4 correspond to generating proposals using each of the 4 proposal methods of Section \ref{sec:iterator}: RAND, EXPLOIT, THOMPSON (with $k=5$ and $p=0.75$), and NN.  The fifth method is a combination of these, labeled ``EVERYTHING'', with the number drawn from each method set to: 4, 1, 9, and 4, respectively - and using the rejection sampling approach for RAND.  We run the system for 26 rounds for each method, repeated 90 times with different random initialization and labeling, and report results averaged across the 90 runs.  For each run we randomly hold out 2000 designs (with stratified sampling) to use as test cases for measuring test accuracy of the user preference models.  For the accuracy metric, we report Area Under the Curve (AUC) as the data is imbalanced.

For the user preference model class, we use a support vector machine classifier per embedding with an RBF kernel, fixing the error weight $C$ to 100 and setting the kernel width $\sigma^2$ to half the 10th percentile of all pairwise distances for that embedding.  The per-embedding classifiers are then combined via a logistic regression classifier.

\subsubsection{Results}

\begin{figure}[htb]
	\centering
	\begin{subfigure}[t]{0.4\textwidth} 
		\includegraphics[width=\textwidth]{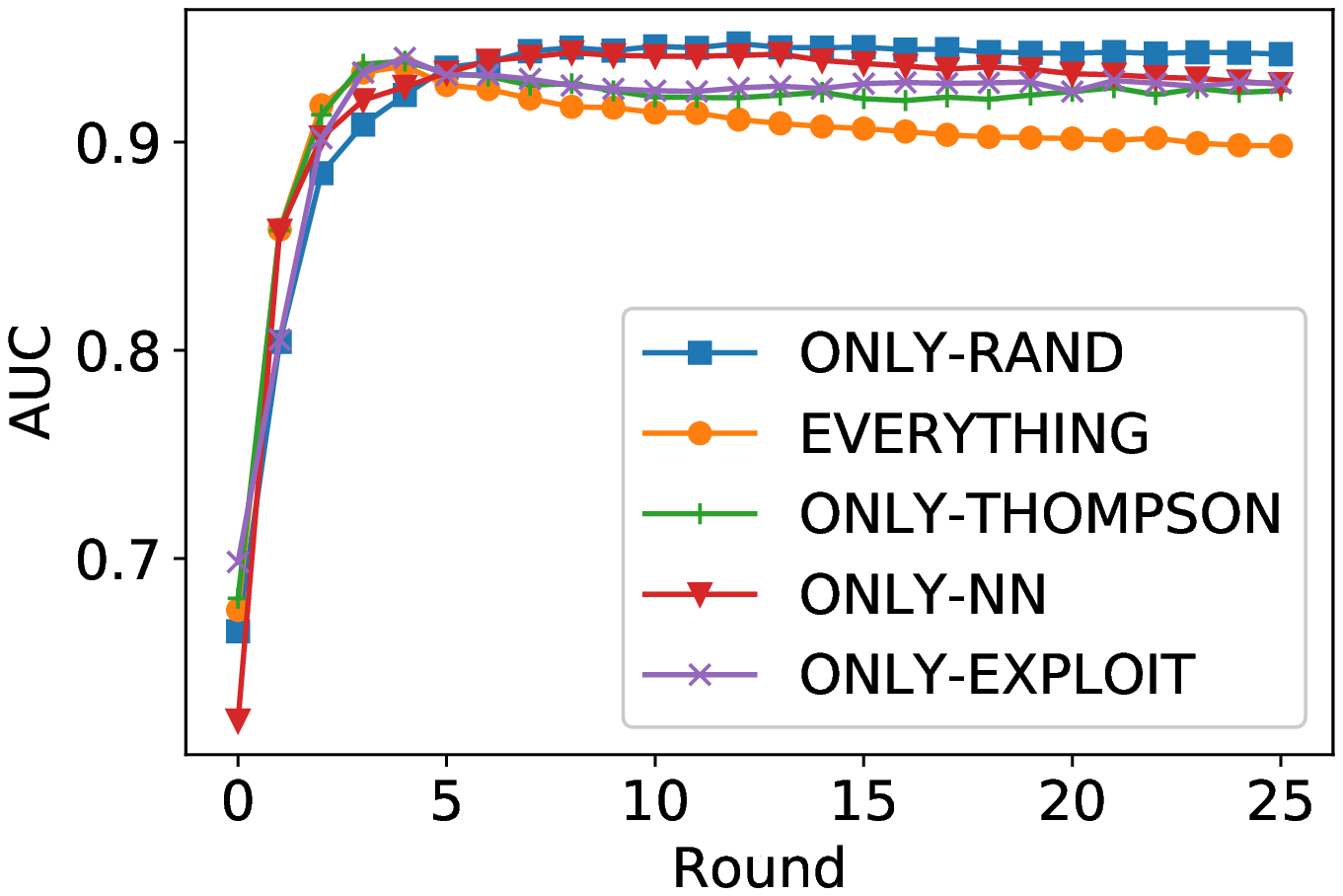}
		\caption{RED} 
	\end{subfigure}
	\begin{subfigure}[t]{0.4\textwidth} 
		\includegraphics[width=\textwidth]{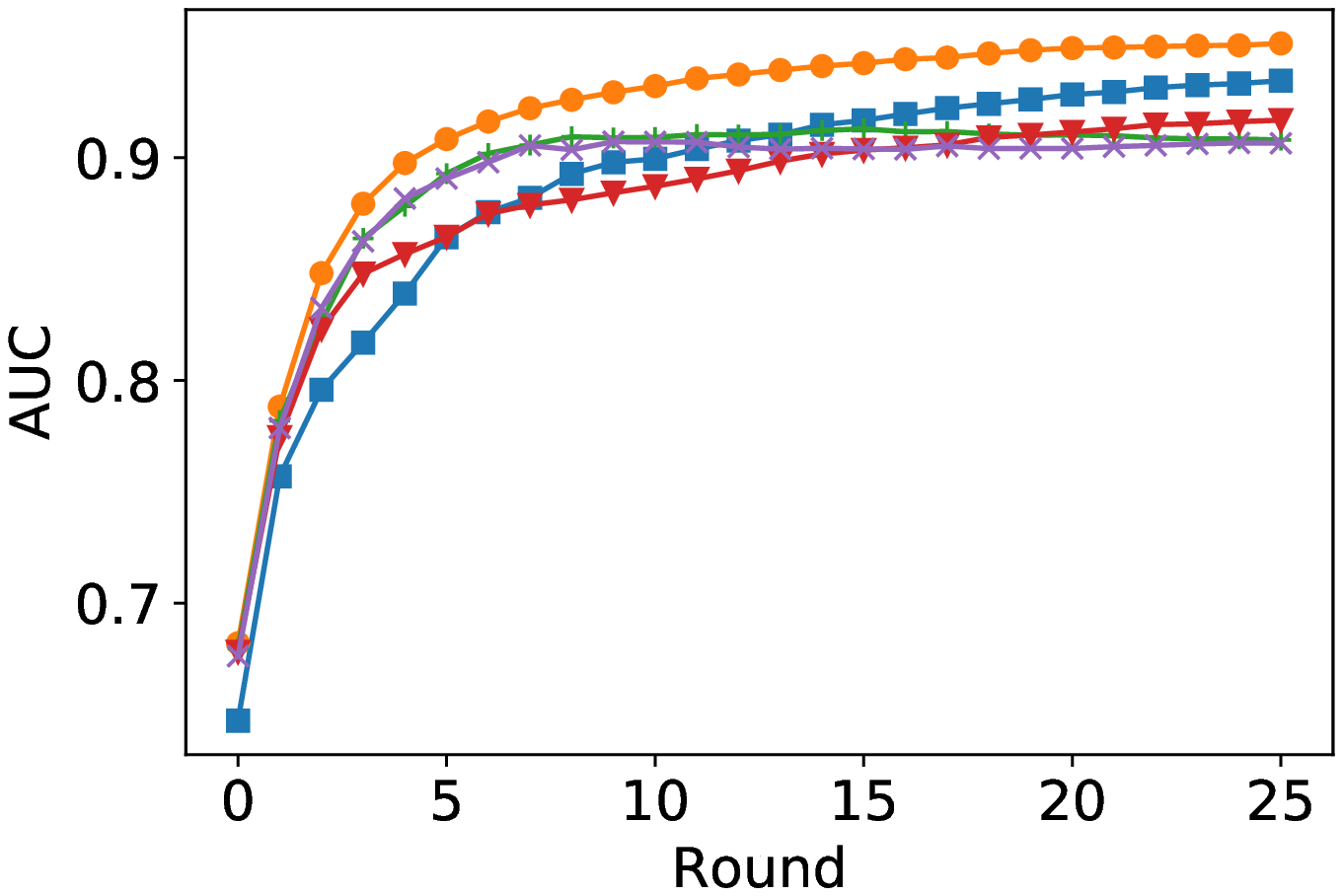}
		\caption{THIN} 
	\end{subfigure}
	\caption{Test AUC per round comparison for two of automated tasks with interactive explorer system.  \label{fig:auto-auc} }
\end{figure}

\begin{figure}[htb]
	\centering
	\begin{subfigure}[t]{0.4\textwidth} 
		\includegraphics[width=\textwidth]{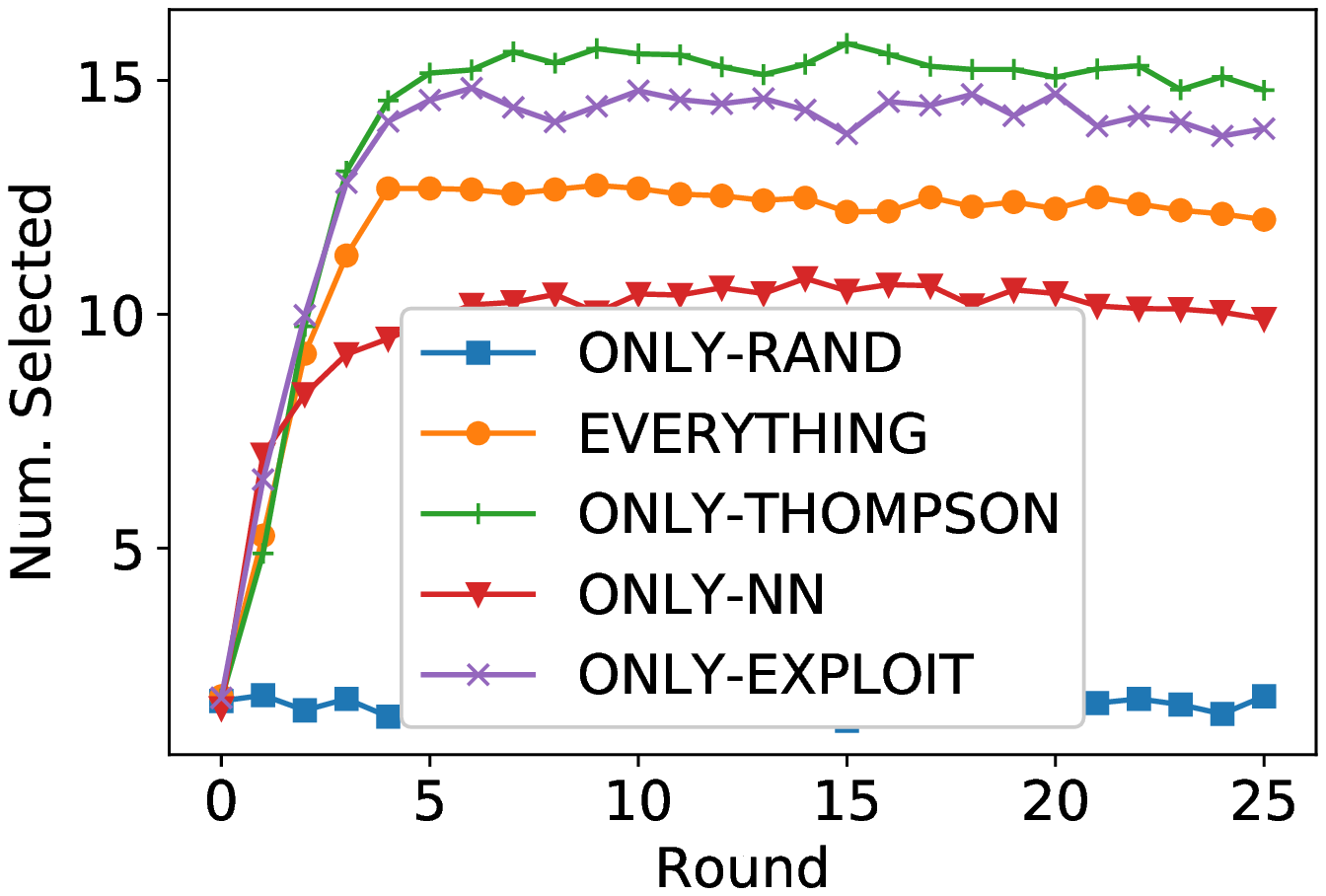}
		\caption{RED} 
	\end{subfigure}
	\begin{subfigure}[t]{0.4\textwidth} 
		\includegraphics[width=\textwidth]{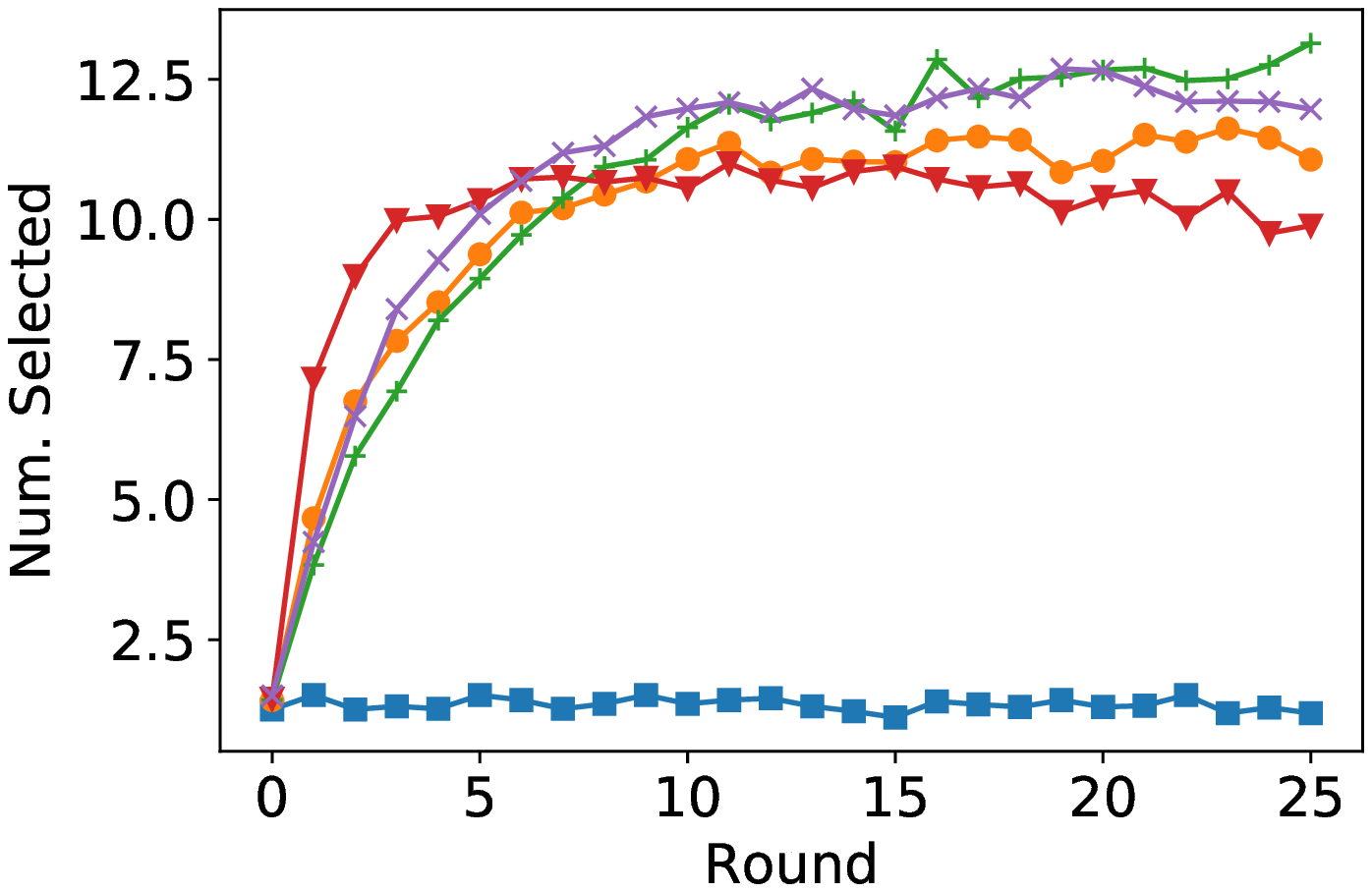}
		\caption{THIN} 
	\end{subfigure}
	\caption{Number selected per round comparison  for two of automated tasks with interactive explorer system. \label{fig:auto-nselect} }
\end{figure}

For simplicity, because results were similar, we only present the results of the automated tasks ``RED'' and ``THIN'' in Figures \ref{fig:auto-auc} and \ref{fig:auto-nselect}, which depict test AUC per iteration and number of user-selected designs per iteration, respectively, averaged over the 90 random runs.  Generally all methods can achieve high AUC and learn the target concept fairly quickly - but the combination method (EVERYTHING) achieves among the top AUC per round in most cases, standing out in particular for the THIN task which may be the most complex task.   We note that for the RED task those methods that most quickly achieve high AUC see subsequent degradation in model accuracy - we suspect this may be due to a resulting early imbalance of positive cases seen in the training data and propose to investigate rebalancing for training in future work.


In terms of number of user-selected proposals per iteration, we see that different methods yield the fastest increase and most total selections for different tasks, except for RAND which is never able to generate a significant number of relevant proposals. Most methods quickly provide more and more relevant proposals, while still allowing some exploration.  Note also that, unlike reinforcement learning, the goal here is not just to get the most selections, but also to incorporate a significant amount of exploration to try and inspire the user to discover new creative ideas or directions.  For this reason we feel if there are not some rejections each round then not enough exploration is included, and the combined approach, EVERYTHING, seems to typically provide a large number of selected designs, without overdoing it.



\subsection{User Study}\label{sec:user-study}
We performed a user study of the proposed AI design assistant with 13 volunteers (7 male, 6 female), to evaluate its use in a simulated real, initial design process.  These were non-expert volunteers since a key use-case of our tool is to assist novice designers or even non-designer business users, such as marketers, needing some quick seed ideas.  First asking them to consider how they would design a perfume bottle on their own, we then explained and showed them each component of the AI design framework.  We then randomly selected one or more design tasks for each and asked them to try out the interactive Iterator component for the design task, and provide feedback. 

Specifically, we asked the following 5 questions about their experience compared to without using the framework.  On a scale of 1 to 5 (5 being the most, and 1 being not at all): \textbf{(Q1)} ``How useful do you think the whole suite of tools would be to help in your design process for coming up with a new, creative design?''; \textbf{(Q2)} ``How much did it help you discover candidate designs related to the task?''; and \textbf{(Q3)} ``How much did it help you explore the wide space of designs and stimulate your creativity?''.  Questions 4 and 5 were free response questions: \textbf{(Q4)}: ``What did you like most about the tool?'' and \textbf{(Q5)}: ``What would you most want to improve about the tool?''.

There were 6 design tasks given as short text descriptions, which mirrored the ambiguous briefings usually provided to designers, meant to capture abstract feelings or concepts.  An example of 2 design tasks were ``The smell is sweet, fruity, girly, and flirty'' and ``Funky, quirky, unique, - they want to stand out from the crowd''.

\subsubsection{Results}
Overall we received positive responses\footnote{Complete responses, question and task details and results, and code will be made available upon publication.}, users seemingly pleasantly surprised and excited about the capabilities offered and the ability of the Iterator to quickly suggest related designs.  The mean and standard deviation of scores for the first 3 questions are given in Table \ref{tab:response-scores}, and percentage of ratings $\ge 4$ in Table  \ref{tab:response-scores-2}, suggesting users did find the proposed system useful.

\begin{table}[htb]
\small
\centering
\begin{tabular}{c|c|c}
 Q1 & Q2 & Q3 \\\hline
 4.19 $\pm$ 0.38 & 4.31 $\pm$ 0.60 & 4.00 $\pm$ 0.61 \\\hline
\end{tabular}
\caption{Mean $\pm$ std. dev. of user ratings per question}
\label{tab:response-scores}
\vspace{-2mm}
\end{table}

\begin{table}[htb]
\small
\centering
\begin{tabular}{c|c|c}
 Q1 & Q2 & Q3 \\\hline
 100 & 92.3 & 76.9 \\\hline
\end{tabular}
\caption{Percent of user ratings $\ge$ 4 per question}
\label{tab:response-scores-2}
\vspace{-2mm}
\end{table}

\begin{figure}[htb]
	\centering
	\begin{subfigure}[t]{0.3\textwidth} 
		\includegraphics[width=\textwidth]{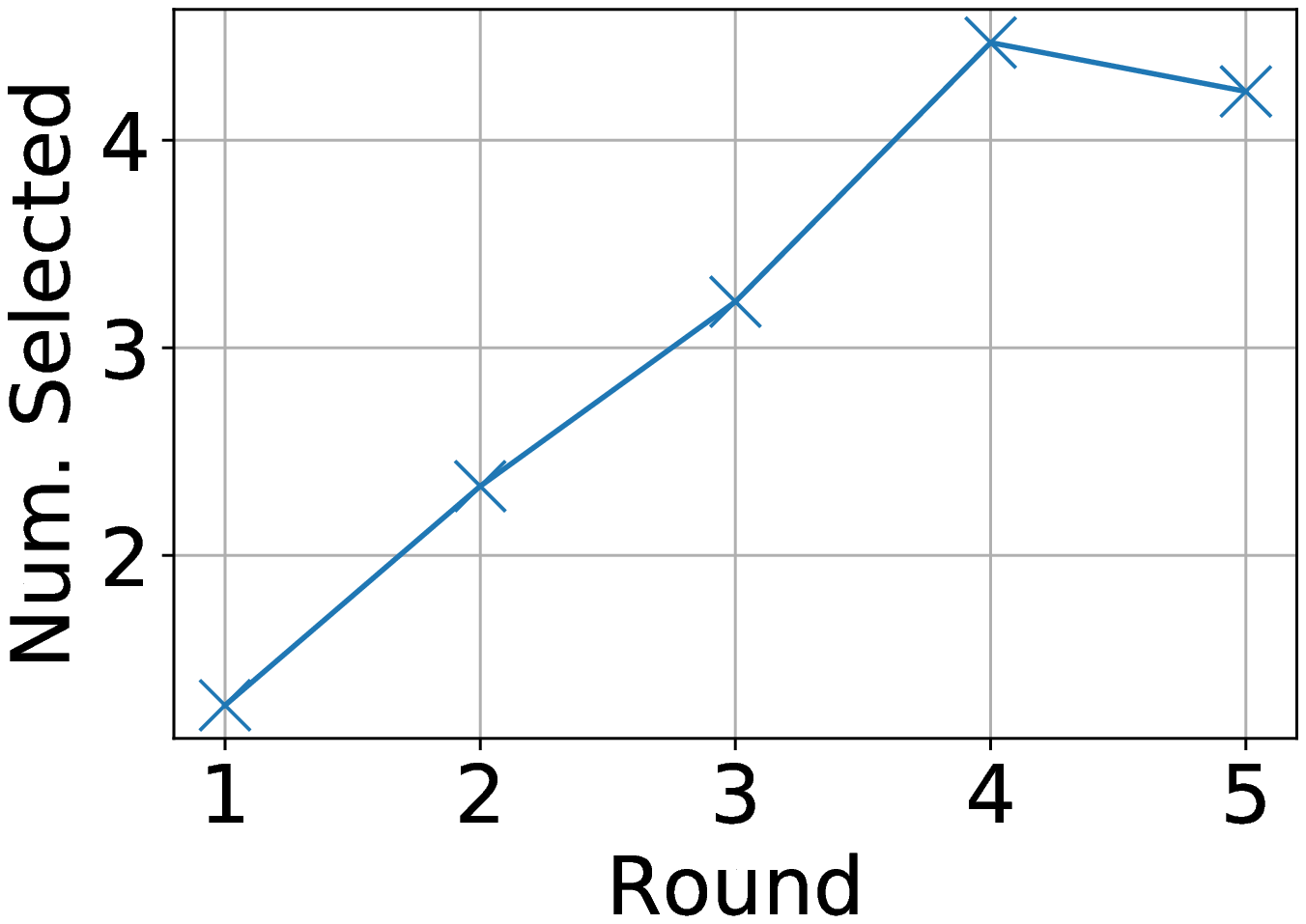}
		\caption{Num. select} 
	\end{subfigure}
	\begin{subfigure}[t]{0.3\textwidth} 
		\includegraphics[width=\textwidth]{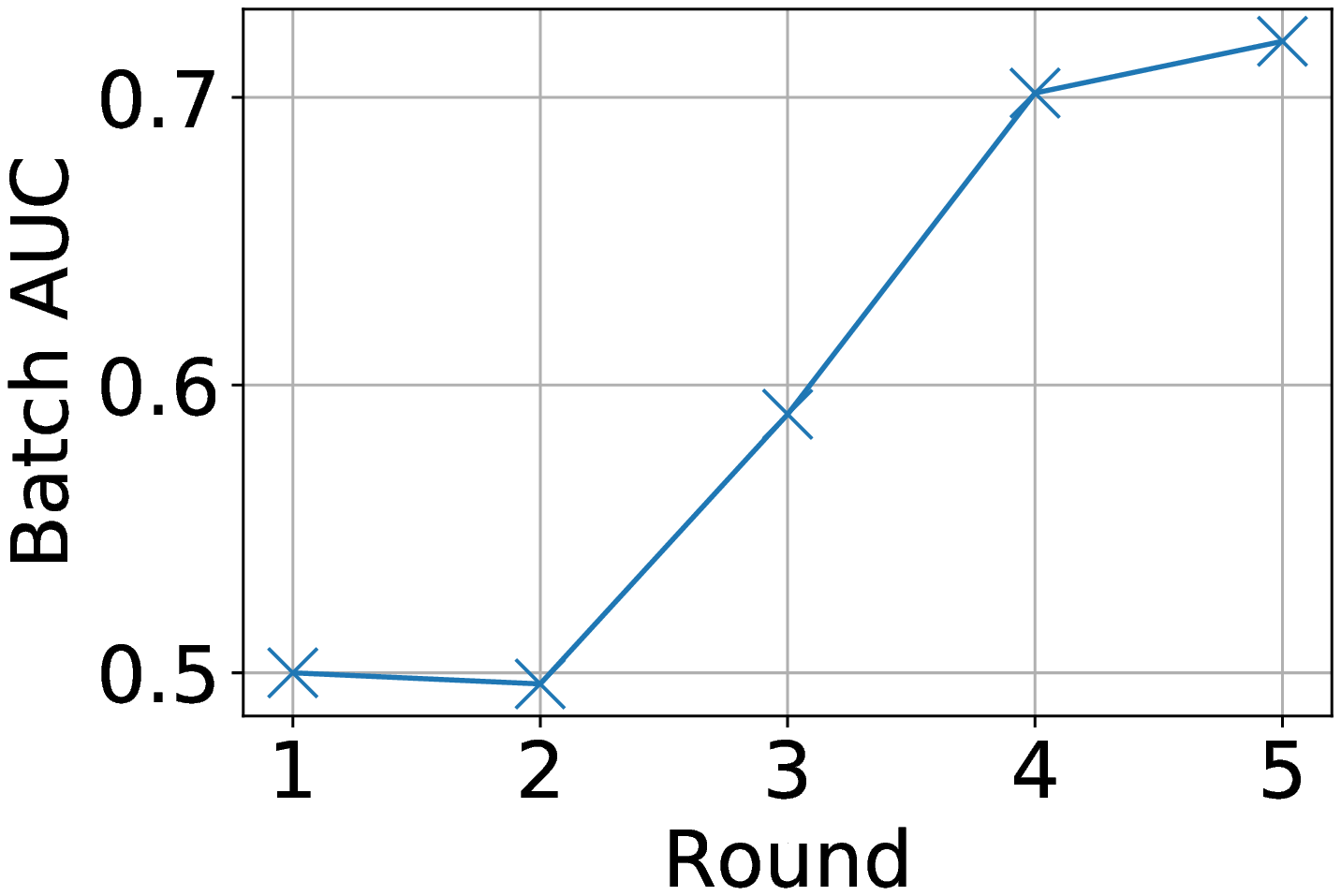}
		\caption{AUC} 
	\end{subfigure}
	\begin{subfigure}[t]{0.3\textwidth} 
		\includegraphics[width=\textwidth]{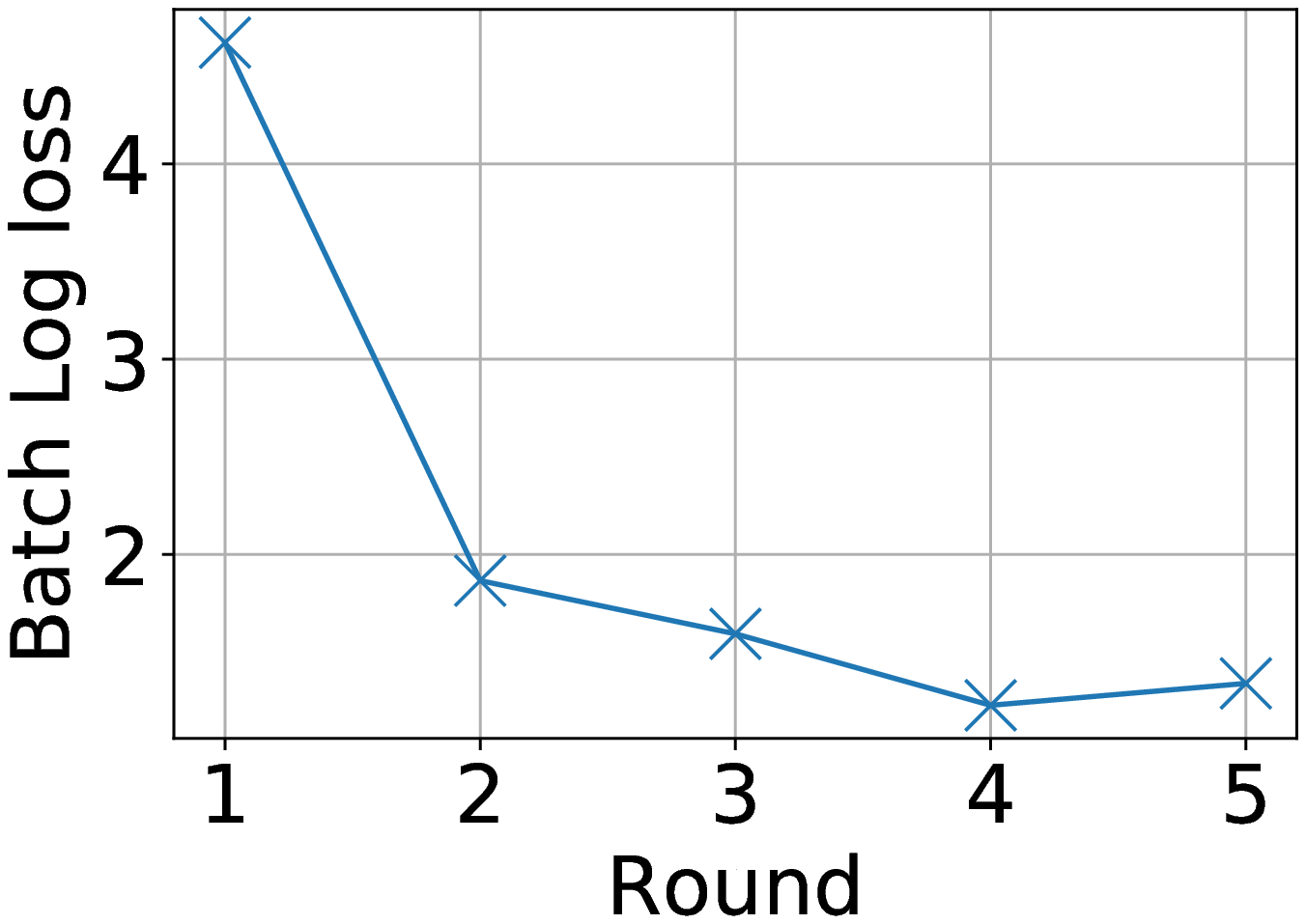}
		\caption{Log loss} 
	\end{subfigure}
	\caption{Number selected, AUC, and log-loss per round. \label{fig:ustudy-res} }
\end{figure}

Since the ground truth concepts were only in the users' minds, we could not hold out a test set to evaluate performance, so instead we report average AUC, log-loss, and number selected per round across the batch of 18 proposal, averaged across 18 runs (some users tried multiple design tasks), in Figure \ref{fig:ustudy-res}.  We show results for the first 5 rounds, since we only enforced users to go through 5 rounds.  As in the automated study, we see the system is able to quickly learn the user's preference and increase selections.

From analyzing the detailed responses, some themes and insights emerged.  Users felt the sketch-to-design and style transfer components could be very helpful and save a lot of time.  For the Iterator, users liked how easy-to-use it was - the aspect of being able to see many different designs and simply select what they liked.  For the most part users felt it could hone-in on their design concepts, and felt the exploration aspect was useful and in some cases provided interesting or unexpected ideas.  However, users sometimes felt their were too many similar designs presented and wanted to see more variation around their target concept, as well as an ability to see past selections and use them to derive further variations.  Having more capabilities to control the exploration was a key desire - such as the ability to fix the color theme, or take aspects of designs they liked, like shape or color only, or vary certain aspects of selected designs.

\subsection{Additional related work}\label{sect:literature}
In addition to co-creativity, the presented framework is related to and leverages several areas of work, which we briefly summarize here.  A number of packaging design tools exist that allow a user to create 2D or 3D forms of structural and graphic designs. These include ESKO ArtiosCAD \cite{eskoArtiosCAD}, ESKO DeskPack \cite{eskoDeskPack}, Adobe Illustrator \cite{adobeIllustrator}, Adobe Photoshop \cite{adobephotoshop} and Creative Edge iC3D \cite{creativeEdgeic3d} . These typically require extensive training and hands-on learning. There also exist commercial platforms that match designers according to users' needs \cite{99designs,fiverr}.  

Recent advances in AI techniques such as Generative Adversarial Networks (GANs), neural style transfers and variational autoencoders (VAEs) have enabled explorations of creativity \cite{creativeAI.net} in domains ranging from music \cite{10.1007/978-3-319-55750-2_8} and arts \cite{ha2018a,DBLP:journals/corr/KazakciCK16} to fashion \cite{2018arXiv180701182D,2018arXiv180703133N}.  A key challenge that remains to be addressed is not generating random designs, but how we enable a human in-the-loop to create a useful design with active control, which we attempt with our framework.

The design space exploration aspect of our framework, i.e., trying to find designs matching user intent, is somewhat similar to visual/semantic search \cite{zhou2017recent,googleImages}.  However, unlike in search, the designer may not know or be able to articulate the target design up-front, and may further want to build on top of initial concepts. Our system instead allows a user to interact with it and learns about the user's preference based on explicit feedback, and also presents creative suggestions for inspiration that may be outside of what the user originally has in mind.

The interactive learning and design proposal components of our framework is related to reinforcement learning (RL)  \cite{sutton2018reinforcement}.  However,  unlike in RL, the user may be focused on creative exploration around a design goal as opposed to choosing a single best set of designs, and our goal is to help the user discover and come across new design ideas. Similarly with respect to Iterator, active learning research \cite{Wang:2012:ATS:2333112.2333120,Rubens2015} could be used to learn an underlying design concept from as few as possible user labels. However, active learning does not have the concept of exploiting which is essential for the human user to get an immediate benefit from using the system.

Related to our Iterator, there is a large body of work on evolutionary computation to evolve designs \cite{singh2012towards}, including interactive genetic algorithms, e.g., for clothing \cite{kim2000application} and bridge design \cite{buelow2008}.  However, unlike our Iterator, these approaches are bound by the random populations that evolve (i.e., the user simply evaluates evolved examples from a given population), require structured specifications and generative procedures to enable the evolutionary generation (which can be difficult to obtain or result in less-realistic design realizations), and do not learn a re-usable user preference model.  We take a different, learning-based approach with our iterator that is also capable of efficiently exploring existing designs and learning an explicit user preference model which has additional applications, such as filtering and scoring new designs directly.  Incorporating evolutionary computation approaches in our framework for further design evolution could be an interesting area of future work. 

\subsection{Conclusion}
We proposed an interactive AI framework for design, with a realization and implemented application for package design, to speed-up the design process, make it easier for both designers and non-expert users, and enhance creativity and exploration of the design space. It consists of 3 main components: Creator, for quickly creating design realizations and variations and automatically generating designs; Evaluator, for evaluating designs along different modalities of interest, like customer impact and aesthetics, and to help guide design exploration; and Iterator, for interactively exploring the design space and learning the user's current design preference.  We constructed a perfume bottle data set to carry out a proof of concept.  We showed results for a number of Creator components including design generation, sketch and color to design realization, and style transfer. Additionally we implemented and tested a complete realization of the Iterator system and demonstrated it is capable of quickly learning the user preference and generating relevant design proposals.

There are several areas of future work.  One is to test incorporating more modalities, e.g., sales and consumer impact, design cost, etc.  Another important area of work is developing new metrics to evaluate the success of the proposals that specifically incorporate measurements of exploration, creativity, and surprise.  Finally, an important future step is enhancing the framework based on the feedback, e.g., adding the capability for user-controlled design variation in Iterator, and integrating the different components into one tool.

\end{document}